\documentclass[manuscript]{aastex631}
\usepackage{amsmath, amssymb, mathtools, bm, multirow, diagbox, hyperref}

\begin{document}
\title{The Alpha-Proton Differential Flow in the Alfv\'enic Young Solar Wind: From Sub-Alfv\'enic to Super-Alfv\'enic}

\author[0000-0002-8234-6480]{Hao Ran}
\affiliation{State Key Laboratory of Space Weather, National Space Science Center, Chinese Academy of Sciences,
Beijing 100190, People's Republic of China; liuxying@swl.ac.cn}
\affiliation{University of Chinese Academy of Sciences, Beijing 100049, People's Republic of China}

\author[0000-0002-3483-5909]{Ying D. Liu}
\affiliation{State Key Laboratory of Space Weather, National Space Science Center, Chinese Academy of Sciences,
Beijing 100190, People's Republic of China; liuxying@swl.ac.cn}
\affiliation{University of Chinese Academy of Sciences, Beijing 100049, People's Republic of China}

\author{Chong Chen}
\affiliation{School of Microelectronics and Physics, Hunan University of Technology and Business, Changsha 410205, People's Republic of China}

\author[0000-0002-3808-3580]{Parisa Mostafavi}
\affiliation{Johns Hopkins Applied Physics Laboratory, Laurel, MD 20723, USA}

\begin{abstract}

Data obtained from Parker Solar Probe (PSP) since 2021 April have shown the first in situ observation
of the solar corona, where the solar wind is formed and accelerated.
Here we investigate the alpha-proton differential flow and its characteristics
across the critical Alfv\'en surface (CAS) using data from PSP during encounters 8-10 and 12-13.
We first show the positive correlation between the alpha-proton differential velocity and the bulk solar wind speed at PSP encounter distances.
Then we explore how the characteristics of the differential flow vary across the CAS and how they are affected by Alfv\'enic fluctuations including switchbacks.
We find that the differential velocity below the CAS is generally smaller than that above the CAS, and the local Alfv\'en speed well limits the differential speed both above and below the CAS.
The deviations from the alignment between the differential velocity and the local magnetic field vector are accompanied
by large-amplitude Alfv\'enic fluctuations and decreases in the differential speed.
Moreover, we observe that $V_{\alpha p}$ increases from $M_A < 1$ to $M_A \simeq 2$ and then starts to decrease, 
which suggests that alphas may remain preferentially accelerated well above the CAS.
Our results also reveal that in the sub-Alfv\'enic solar wind
both protons and alphas show a strong correlation between their velocity fluctuations and magnetic field fluctuations,
with a weaker correlation for alphas.
In contrast, in the super-Alfv\'enic regime the correlation remains high for protons, but is reduced for alphas.

\end{abstract}


\section{Introduction} \label{sec:intro}

The two main ion components of the solar wind, protons and alpha particles, 
are different in terms of their velocity and temperature
\citep[e.g., ][]{marsch_solar_1982,steinberg_differential_1996,neugebauer_ulysses_1996,kasper_solar_2007,kasper_evolution_2012,mostafavi_alphaproton_2022}.
Despite being heavier, alpha particles tend to travel faster than protons, which generates a differential streaming between them.
This phenomenon was first reported by \cite{robbins_helium_1970}, and remains an unsolved puzzle.
Several mechanisms have been proposed to explain the cause of the differential streaming.
For example, \cite{isenberg_preferential_2009} proposed a cyclotron-resonant Fermi process that is capable of providing 
the preferential acceleration of minor ions, \cite{verscharen_parallel-propagating_2013} discussed the role of the parallel Alfv\'enic 
drift instabilities in the formation and evolution of the differential streaming, and 
\cite{chandran_alfven-wave_2010} claimed that stochastic heating caused by low-frequency Alfv\'en-wave turbulence 
also contributes to the preferential heating of minor ions.
According to \cite{Hu1999resonant}, the preferential acceleration and heating of the alpha particles close to the Sun
are both due to the resonant cyclotron interaction.

Before the Parker Solar Probe \citep[PSP; ][]{fox_solar_2016} era, the differential flow was studied with data from 0.3 to 5.5 au.
Some common features of the differential velocity emerged in this range of distances.
For example, using data from Helios 1 and 2, \cite{marsch_solar_1982} reported a clear tendency of the differential 
speed increasing with the solar wind speed. The same conclusion was drawn by \cite{steinberg_differential_1996}
and \cite{durovcova_evolution_2017} with Wind observations.
In addition, previous observations suggested that the differential speed is of the order of, but generally lower than, 
the local Alfv\'en speed \citep[e.g.,][]{marsch_solar_1982,neugebauer_ulysses_1996,berger_systematic_2011,durovcova_evolution_2017}.
One possible interpretation was given by \cite{gary_electromagnetic_2000} that a sufficiently fast differential flow can lead to alpha-proton instabilities 
that could drive wave-particle scattering.
This scattering reduces the differential speed and thus limits it to be lower than the local Alfv\'en speed.
Another interesting feature is the alignment between the differential velocity and the local magnetic field vector \citep[e.g.,][]{marsch_solar_1982,yamauchi_differential_2004, matteini_ion_2015}.
\cite{yamauchi_differential_2004}, for example, revealed a high correlation between the 
differential velocity and the magnetic field vector both inside and 
outside pressure balance structures with Ulysses observations.

The PSP mission, launched in 2018, has enabled us to investigate the differential flow at unprecedented proximity to the Sun \citep{mcmanus_density_2022,mostafavi_alphaproton_2022}.
\cite{mostafavi_alphaproton_2022} confirmed the positive correlation between the differential velocity and the bulk solar wind speed,
the dependence of the differential flow on heliocentric distance, 
and the limitation of the differential speed by the local Alfv\'en speed inside 0.3 AU, using PSP data from encounters 3-7.
However, sub-Alfv\'enic solar wind measurements are not incorporated into their study,
and the impacts of Alfv\'enic fluctuations on the differential flow are not discussed.
\cite{ofman_observations_2023} 
studied the nonlinear evolution of ion kinetic instabilities and the transfer of energy between ions and waves in the sub-Alfv\'enic regime, 
which suggests that the locally produced ion instabilities contribute to the anisotropic heating of alphas.
One of the striking observations of PSP is the common presence of Alfv\'enic switchbacks in the young solar wind,
which are characterized by a deflection in the magnetic field direction and an increase in the radial velocity \citep{bale_highly_2019, kasper_alfvenic_2019}.
\cite{mcmanus_density_2022} examined the 3D nature of the velocity fluctuations of alphas and protons in switchbacks 
using PSP observations from encounters 3 and 4. 
They revealed that the velocity of alphas may increase, remain constant, or decrease in switchbacks 
depending on whether the differential speed is smaller, equal to, or larger than the local wave speed.
Here our study includes the behaviors of alphas and protons inside switchbacks at closer distances to the Sun.
Another important achievement of PSP is the crossing of the critical Alfv\'en surface (CAS) for the first time in 2021 April \citep{kasper2021parker},
which happened at a surprisingly large distance ($\sim$ 19 $R_S$).
\cite{Liu_2023} suggested that the sub-Alfv\'enic solar wind detected by PSP is a special type of wind emanating from boundaries 
inside a coronal hole, termed a low Mach-number boundary layer (LMBL).
An LMBL is associated with an enhanced Alfv\'en radius, which explains the crossing of the CAS at $\sim$ 19 $R_S$.

In this paper, we focus on the characteristics of the alpha-proton differential flow across the CAS and how they are affected by Alfv\'enic fluctuations including switchbacks.
Given the magnetic nature of switchbacks and the unprecedented measurements of the sub-Alfv\'enic solar wind,
it is of great interest to investigate the above mentioned properties of the differential flow at PSP encounter distances,
especially in the new domain below the CAS,
and to explore the effects of Alfv\'enic fluctuations on the differential flow.
Specifically, this work aims to address the following questions:
(i) How does the alpha-proton differential flow vary with the solar wind speed in the most recent orbits of PSP?
(ii) How do the characteristics of the differential flow vary when the solar wind changes from sub-Alfv\'enic to super-Alfv\'enic (i.e., across the CAS)?
(iii) How do Alfv\'enic fluctuations including switchbacks affect the differential flow?
This work is organized as follows. 
In Section \ref{sec:data}, we present our analysis and results.
We summarize our conclusions in Section \ref{sec:con}.

\section{Observations and Results} \label{sec:data}
Data used in this study are from the FIELDS \citep{bale_fields_2016} instrument suite and the SWEAP \citep{kasper_solar_2016} package aboard PSP.
The level 2 product from FIELDS provides us the RTN magnetic field data.
The ion data are from the level 3 moment products from SPAN-I, which is an electrostatic analyzer on the ram side of PSP \citep{livi_solar_2022}.
We consider an interval of observations to be reliable if the cores of the ions (protons and alphas) are within the field of view and energy range of SPAN-I.
Further, transient events such as coronal mass ejections (CMEs) or heliospheric current sheet (HCS) crossings in the selected intervals 
are excluded.
The details of the data selection and curation in this study are given in the Appendix.
Densities measured by SPAN-I are derived from partial moments of the particle velocity distributions, 
which may not be accurate as the full ion distribution is not captured by SPAN-I. 
Therefore, we adopt the electron density from quasi-thermal noise (QTN) spectroscopy to represent the
plasma density \citep{moncuquet_first_2020}.
The local Alfv\'en speed is thus computed as
$V_{A} = B / \sqrt{\mu_0 n_e m_p}$,
where $B$ is the local magnetic field strength, $\mu_0$ is the permeability of vacuum,
$n_e$ is the electron density, and $m_p$ is the proton mass.
Use of electron density as a representation of the plasma density may introduce an overestimate of the local Alfv\'en speed by 2\% to 5\% \citep{mostafavi_alphaproton_2022}.
We use PSP data from encounters 8-10 and encounters 12-13, during which sub-Alfv\'{e}nic solar wind measurements are 
reported \citep{Jiao_2024}.
We exclude encounter 11 due to the absence of the QTN density for this encounter.
The data are linearly interpolated to 2 s resolution.
All vector data are in the RTN coordinate system.

The magnitude of the differential velocity is calculated as:
\begin{equation}
 |\bm{V_{\alpha p}}| = \mathrm{sign}(|\bm{V_{\alpha}}| - |\bm{V_p}|)(|\bm{V_{\alpha}} - \bm{V_p}|).
\end{equation}
This derivation takes the direction of both the protons and alphas into consideration \citep[e.g.,][]{steinberg_differential_1996, berger_systematic_2011,durovcova_evolution_2017,mostafavi_alphaproton_2022}.
The method for calculating the deflection angle (i.e., $\theta$) of the magnetic field is adopted from \cite{Liu_2023}.
The calculation is capable of indicating the deflection direction of the magnetic field.
The angle $\theta$ varies from $-\pi$ to $\pi$, with a negative one corresponding to a clockwise deflection and a positive one corresponding to a counter-clockwise deflection when viewed from the north.
A value of $-\pi$ or $\pi$ implies a completely reversed magnetic field.
We include all possible deflection angles instead of only those larger than $90^{\circ}$ \citep[e.g.,][]{wit_switchbacks_2020,Liu_2023}.
\cite{Liu_2023} showed that with this derivation of the deflection angle the enhancement in the proton radial velocity can be written as 
$\delta V_{pR} / V_A = 1 - \cos\theta$, which well explains the one-sided nature of the velocity spikes observed
within switchbacks \citep[e.g., ][]{gosling_one-sided_2009, horbury_short_2018,kasper_alfvenic_2019}.
Readers are directed to \cite{Liu_2023} for a detailed description of the derivation of the deflection angle and 
its relationship with the velocity enhancement.
Also, following \cite{Liu_2023}, we derive the variation of the radial velocity of the particles $\delta V_R$ by subtracting a low-pass filtered radial velocity $V_{Rf}$ from $V_R$.
The cuttoff frequency of the filter is $2 \times 10^{-4}$ Hz, which is below the break frequency (about $2 \times 10^{-3}$ Hz) between the energy-containing range and inertial range in the power spectrum of solar wind fluctuations
\citep[e.g.,][]{kasper2021parker,Zank_2022}.

\subsection{Overview of Measurements at Encounters 9 and 10}

\begin{figure}
    \centering
    \includegraphics[width=1.0\textwidth, height=1.0\textwidth]{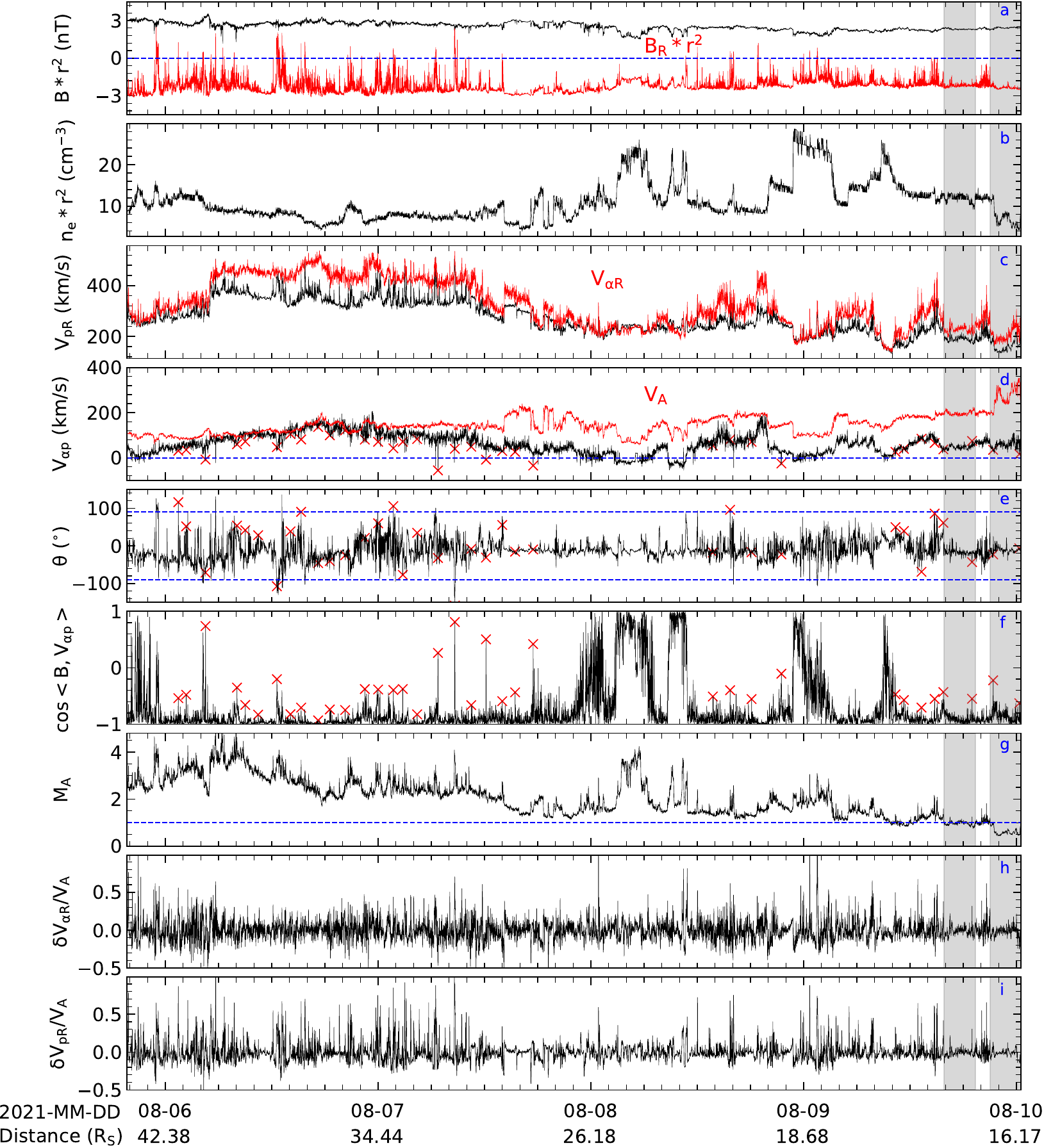}
    \caption{
        PSP measurements at encounter 9 showing a case of the slow solar wind. 
    The shaded areas indicate the sub-Alfv\'enic intervals.
    (a) Normalized (to 1 au values) magnetic field magnitude and radial component (red).
    (b) Normalized electron density (from QTN).
    (c) Radial velocity of protons and alpha particles (red).
    (d) Differential speed and local Alfv\'en speed (red).
    (e) Magnetic field deflection angle.
    (f) Cosine of the angle between magnetic field vector and differential velocity.
    (g) Alfv\'en Mach number.
    (h) Radial velocity change of alpha particles in units of local Alfv\'en speed.
    (i) Radial velocity change of protons in units of local Alfv\'en speed.
    The times of the red crosses are derived by identifying the spikes in panel (f) every 50 data points
    when the differential speed is overall larger than 0.}
    \label{fig:en9}
\end{figure}

We first present an overview of the measurements at encounters 9 and 10, 
which show a slow solar wind case and a fast solar wind case, respectively.
Figure \ref{fig:en9} displays the in situ measurements near the perihelion of encounter 9 made from 
19:40 UT on 2021 August 5 to the beginning of 2021 August 10, during which PSP distances range from 
about 42 $R_S$ to 16 $R_S$.
We select this time range, because it covers a relatively wide range of $V_{\alpha p}$ and contains sub-Alfv\'enic intervals.
The radial velocity of protons in this interval starts from around 300 km/s, then rises up to about 380 km/s
at around 05:00 UT on August 6, and stays at this level until around 12:00 UT on August 7.
After this plateau, the radial velocity of protons drops to around 220 km/s and varies in the range from 170 km/s to 300 km/s.
As can be seen in Figure \ref{fig:en9}(c),
in the faster solar wind the radial speed of alphas is higher than that of protons, 
while in the slower solar wind the two radial speeds are close.
In the sub-Alfv\'enic intervals the solar wind is very slow but presents a relatively obvious $V_{\alpha p}$.
These slow but fast wind like flows are consistent with the theory of \cite{Liu_2023}, who term such sub-Alfv\'enic intervals LMBLs.
An LMBL, characterized by a drop in the Alfv\'en Mach number resulting from its decreased plasma density and relatively low velocity,
 may originate from the peripheral region of a coronal hole.
Figure \ref{fig:en9}(d) shows the relationship between $V_{\alpha p }$ and $V_A$.
The local Alfv\'en speed acts as an upper limit for $V_{\alpha p}$.
For the faster wind from 06:00 UT on 2021 August 6 
to 00:00 UT on 2021 August 7, $V_{\alpha p}$ fluctuates near $V_A$.
The average speed of this interval is 358 km/s.
Figure \ref{fig:en9}(f) displays the cosine value of the angle between the magnetic field vector and the differential velocity.
The two vectors are generally well aligned (i.e., close to -1 or 1) except when $V_{\alpha p}$ is close to 0 and changes sign frequently.
Such frequent changes in the sign of $V_{\alpha p}$ cause the cosine value to fluctuate between -1 and 1, 
which makes the determination of the alignment difficult.
Therefore, we mark the spikes of the curve in Figure \ref{fig:en9}(f) every 50 data points with red crosses
when $V_{\alpha p}$ is overall positive.
The red crosses then represent the times when the alignment deviates.
We also mark those points in Figure \ref{fig:en9}(d) and (e) with crosses too.
The locations of the crosses indicate that the spikes in Figure \ref{fig:en9} (f) are associated with drops in $V_{\alpha p}$ and large deflection angles. 
This implies that the deviations from the alignment are related to switchbacks and the velocity fluctuations within switchbacks.
Figure \ref{fig:en9}(i) shows that the velocity variations of protons are largely one-sided as in \cite{Liu_2023}.
The velocity variations of alphas (Figure \ref{fig:en9}(h)) are similar to those of protons, but appear more two-sided.
This may suggest a weaker modulation of alphas by Alfv\'enic fluctuations.
\cite{mcmanus_density_2022} showed cases where the velocity of alphas may increase, remain constant, or decrease in switchbacks depending on the relationship between the differential speed and the wave speed.
This explains the more two-sided velocity variations of alphas.
We will discuss this result in more details below.

\begin{figure}
    \centering
    \includegraphics[width=1.0\textwidth, height=1.0\textwidth]{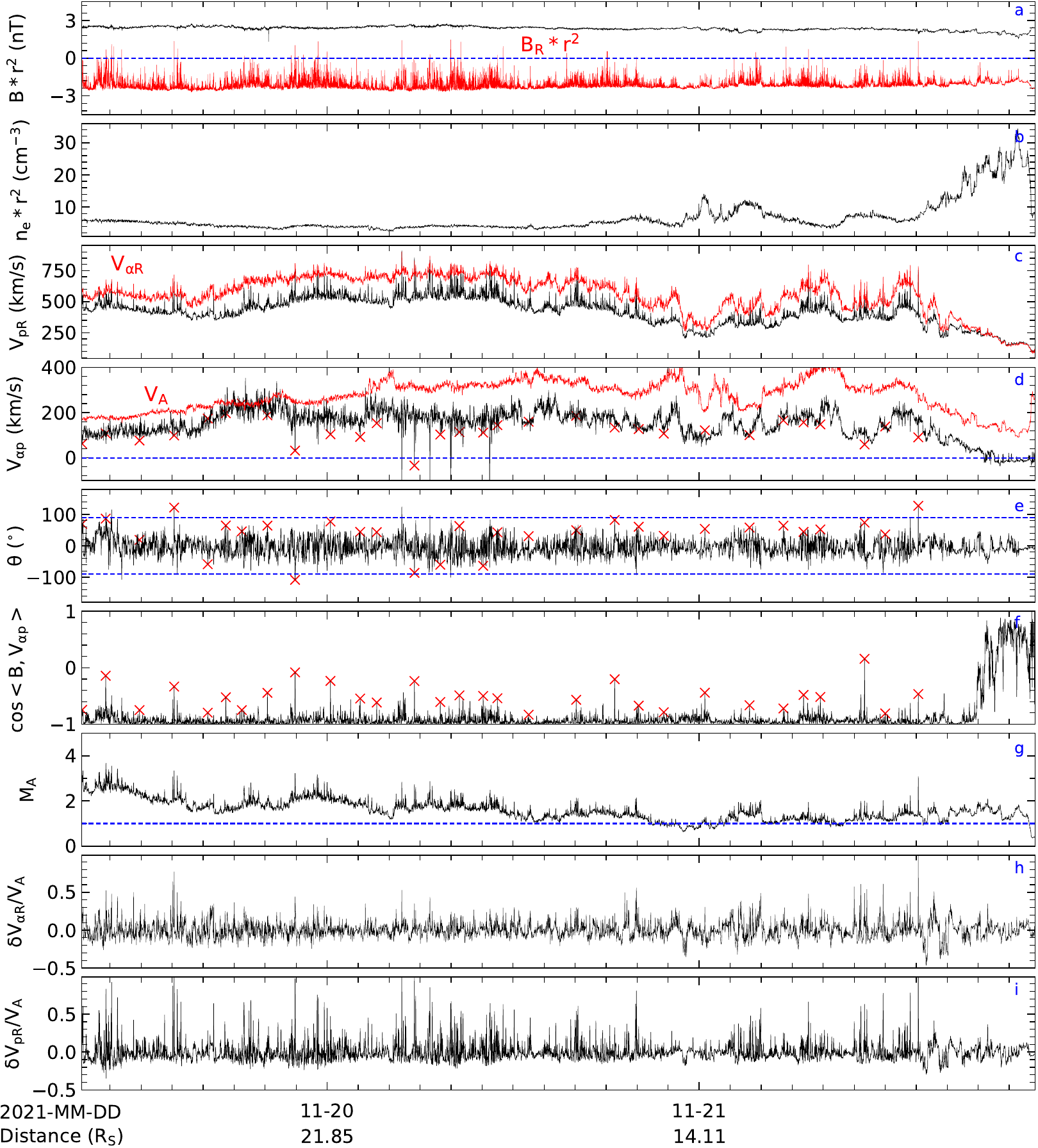}
    \caption{PSP measurements at encounter 10 showing a case of the fast solar wind. Similar to Figure \ref{fig:en9}.}

    \label{fig:en10}
\end{figure}

Figure \ref{fig:en10} presents the measurements at encounter 10 
from 08:10 UT on 2021 November 19 to the end of 2021 November 21.
The reason for selecting this time range is similar as for encounter 9, but here we focus on a fast wind interval.
The radial speed of protons first starts from around 450 km/s, 
and rises to over 500 km/s at the beginning of 2021 November 20.
Our magnetic mapping results (not shown here) suggest that 
PSP was connected to the interior of a large coronal hole during this interval.
The magnetic mapping of the same encounter can also be found in \cite{badman_2023_mapping}.
Then it drops to around 250 km/s at the end of 2021 November 20, and soon rises up again to over 400 km/s.
This dip of the radial speed of protons corresponds to an interval when PSP was surfing around the CAS (See Figure \ref{fig:en10}(g)).
The radial speed of protons finally drops to a very low level after 14:00 UT on 2021 November 21.
As at encounter 9, the radial velocity of alphas is higher than that of protons in the faster solar wind.
The local Alfv\'en speed is an upper limit for $V_{\alpha p}$ in most cases (98.2 \%).
The alignment between the magnetic field vector and the differential velocity is in general well-kept 
when $V_{\alpha p}$ does not change sign frequently.
The locations of the crosses indicate that the deviations from the alignment are accompanied by large deflection angles and
drops in $V_{\alpha p}$.
The velocity variations of alphas (Figure \ref{fig:en10}(h)) again appear more two-sided than those of protons.
It is important to mention that, although PSP was traveling in the vicinity of the CAS,
we have not identified a steady sub-Alfv\'enic interval during this time period.

\subsection{Distributions and Correlations}
\begin{figure}
    \centering
    \includegraphics[width=1.0\textwidth, height=0.7\textwidth]{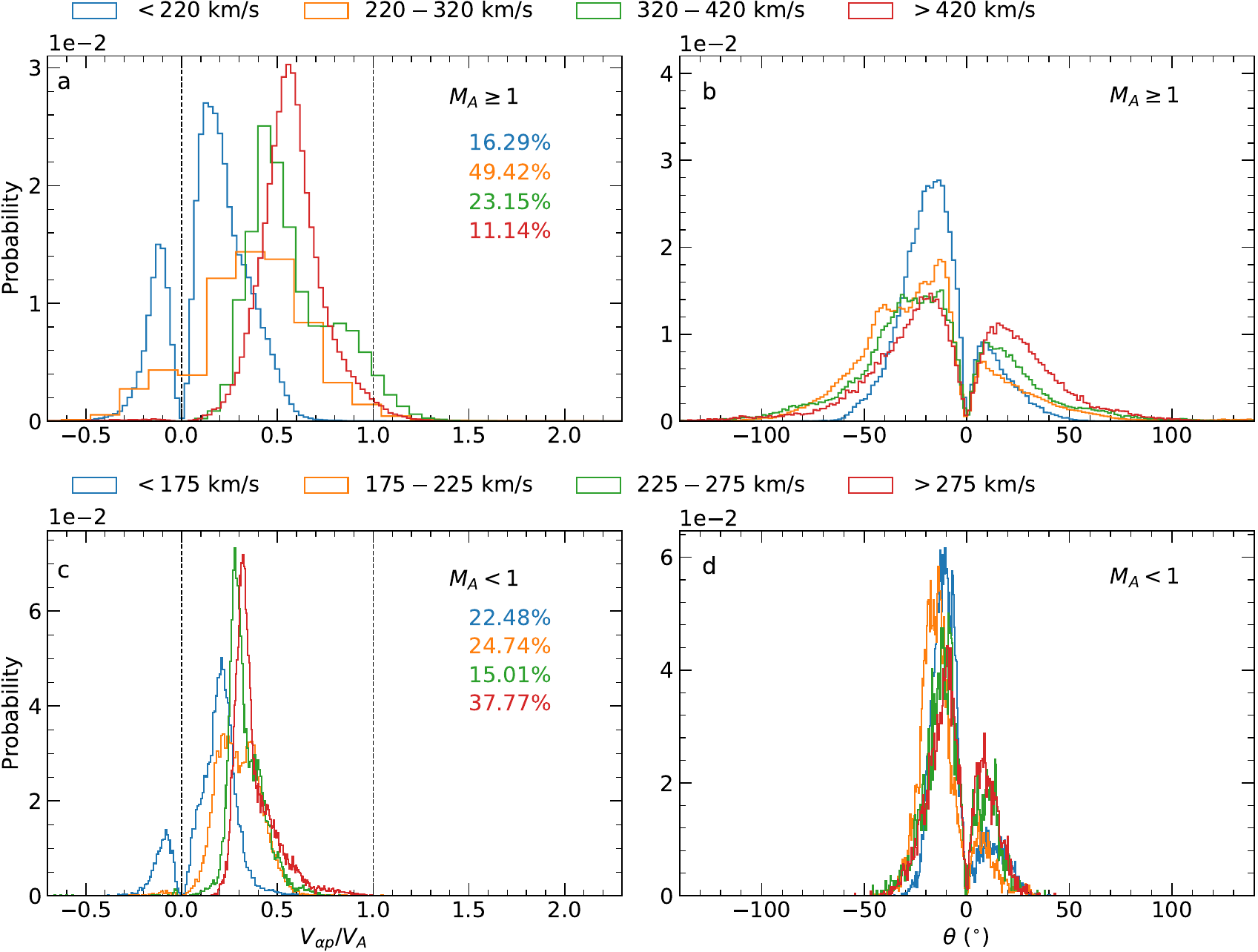}
    \caption{Histograms of $V_{\alpha p} / V_A$ (left) and deflection angle $\theta$ (right) in super-Alfv\'enic (top) and sub-Alfv\'enic (bottom) solar wind. 
    The super-Alfv\'enic and sub-Alfv\'enic solar winds are divided into 4 groups according to their bulk speed at 100 and 50 km/s intervals, respectively.
    The percentages shown in panels (a) and (c) indicate the proportion of each group to the total number of data points.
    }

    \label{fig:hist}
\end{figure}

After showing the typical examples for the slow and fast wind, here we look at the distributions and correlations in the data.
We also include data from encounters 8, 12, and 13 to have enough statistics.
Figure \ref{fig:hist} contains the histograms of $V_{\alpha p} / V_A$ and $\theta$ in the super-Alfv\'enic 
and sub-Alfv\'enic solar wind.
We divide the super-Alfv\'enic (1037029 measurements) and sub-Alfv\'enic (76606 measurements) solar winds into four groups according to their bulk speed at 100 km/s and 50 km/s intervals, respectively.
The groups are chosen such that each group has an acceptable proportion of the data.
As suggested by Figure \ref{fig:hist}(a), for the super-Alfv\'enic solar wind, most of $V_{\alpha p} / V_A$ is between 0 and 1. 
Values lower than 0 (i.e., $V_{p} > V_{\alpha}$) usually occur in the slower solar winds ($V_{p} < 320$ km/s).
Values higher than 1 (i.e., $V_{\alpha p} > V_A$) are rare (only 2.2 \%), and they usually occur in the faster solar winds ($V_{p} > 320$ km/s).
\cite{mcmanus_density_2022} also reported cases when $V_{\alpha p} > V_A$.
They suggested that alphas will move in antiphase with protons inside switchbacks if $V_{\alpha p}$ is larger than the wave speed,
which will result in a spike in the proton velocity and a dip in the alpha velocity.
The reason why $V_{\alpha p}$ could exceed the local Alfv\'en speed, however, is unclear.
The peak in general moves to higher values as the solar wind speed increases. 
These observations are consistent with previous studies \citep[e.g.,][]{marsch_solar_1982, mostafavi_alphaproton_2022}.
Figure \ref{fig:hist}(b) shows the deflection angle distributions of the super-Alfv\'enic solar wind.
The deflection angle is predominantly negative, which indicates a preferential clockwise rotation of the magnetic field \citep[also see ][]{squire_properties_2022,fargette_preferential_2022}.
According to \cite{Liu_2023}, this is due to the clockwise geometry of the background Parker spiral field.
The degree of this preference decreases as the solar wind speed increases, as expected from 
the more radial Parker spiral field in the faster solar wind.
The peak also shifts to higher values as the solar wind speed increases,
which indicates that larger switchbacks tend to appear in the faster solar winds.
In Figure \ref{fig:hist}(c), we rarely observe data points with $V_{\alpha p} / V_A > 1$ in the sub-Alfv\'enic solar wind (less than 0.01 \%), 
and data points with $V_{\alpha p} / V_A < 0$ also exist in the slowest solar winds ($V_p < 175$ km/s).
Akin to that of the super-Alfv\'enic solar wind, the peak of the distributions moves to higher values as the solar wind speed increases.
For the deflection angle distributions of the sub-Alfv\'enic solar wind shown in Figure \ref{fig:hist} (d), the widths of all categories are roughly the same,
and the preference for clockwise rotation still exists.
Equal widths agree with the LMBL nature of the sub-Alfv\'enic intervals.
As explained in \cite{Liu_2023}, a low Alfv\'en
Mach number suppresses the amplitudes of switchbacks, regardless of the solar wind speed.
The preference for clockwise rotation of the magnetic field indicates that the influence brought by 
the clockwise curvature is still valid below the CAS.

\begin{figure}
    \centering 
    \includegraphics[width=1.0\textwidth, height=0.4\textwidth]{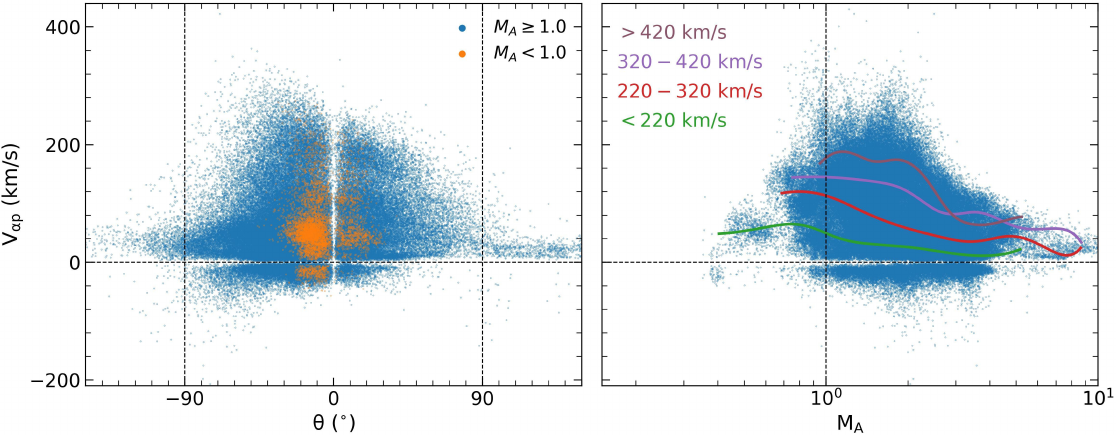}
    \caption{Measurements of $V_{\alpha p}$ as functions of magnetic field deflection angle (left) 
    and radial Alfv\'en Mach number (right). The orange scatter in the left panel refers to the sub-Alfv\'enic measurement.
    The curves shown in the right panel indicate the averages of the alpha-proton differential flow as a function of $M_A$
    for different ranges of the solar wind speeds.}
    \label{fig:Vap_theta_MA}
\end{figure}

Figure \ref{fig:Vap_theta_MA} shows the scatter plots of $V_{\alpha p}$ versus 
$\theta$ and $M_A$.
We notice a rough decreasing trend of $V_{\alpha p}$ with increasing $|\theta|$.
This is consistent with Figure \ref{fig:en9} and Figure \ref{fig:en10} that
large switchbacks are accompanied by drops in $V_{\alpha p}$.
The larger switchbacks may disturb the alignment between the differential velocity and the magnetic field vector in the inertial frame, 
and reduce the magnitude of the differential velocity \citep{matteini_ion_2015, mcmanus_density_2022}.
The sub-Alfv\'enic solar wind is concentrated at small deflection angles as expected,
and the presence of large $V_{\alpha p}$ in the sub-Alfv\'enic wind may indicate the existence of the fast wind like features in the slow
sub-Alfv\'enic solar wind.
The right panel shows $V_{\alpha p}$ as a function of $M_A$.
For a better interpretation, we also plot the averages of $V_{\alpha p}$ for 
each speed category.
The curves show the dependence of the differential speed on both $M_A$ and the solar wind speed.
Firstly, as the solar wind becomes faster at a fixed $M_A$, the differential speed generally increases accordingly.
On the other hand, the increase of $M_A$ can be considered as the acceleration
process of the solar wind from lower to higher altitudes.
We observe that the maximum value of $V_{\alpha p}$ increases until $M_A \simeq 2$, 
which may indicate that a strong preferential acceleration of alphas occurs in this region.
In the slower solar wind ($V_p < 320$ km/s) the preferential acceleration of alphas occurs below the CAS, and then decreases above the CAS. 
In the faster solar wind ($V_p > 320$ km/s) the preferential acceleration region extends out above the CAS.
\cite{kasper_strong_2019} suggested that the preferential heating of alphas, which is highly related to the preferential acceleration of alphas \citep{Kasper2008_HotHelium, Bourouaine_2011},
is predominantly active below the CAS.
According to their discussions, a fraction of the outward propagating waves in the expanding solar wind can
be reflected back toward the Sun, which, if below the CAS, can travel all the way back to the Sun and interact with the outward propagating waves.
Such interactions can lead to wave-reflection-driven turbulence providing the energy for the preferential heating of alphas.
It is unclear why alphas are still preferentially accelerated above the CAS in the fast solar wind.
It may be attributed to the combination of two factors.
On the one hand, the preferential acceleration of alphas is more intense and diverse in the fast solar wind \citep{marsch_solar_1982}.
For example, \cite{isenberg_preferential_2009} proposed an ion cyclotron-resonant Fermi process that 
can perpendicularly heat and accelerate minor ions in a high-speed stream originating from a coronal hole.
On the other hand, Coulomb collisions are practically negligible in the fast solar wind due to its 
high-temperature and low-density conditions \citep{marsch_solar_1982}.
Therefore, it takes longer for Coulomb collisions to thermalize the plasma in the faster solar wind.
This may also contribute to the enhancement of $V_{\alpha p}$ above the CAS in the fast solar wind.
After $M_A \simeq 2$, the differential velocity decreases with increasing $M_A$. 
The decreasing trend of $V_{\alpha p}$ with $M_A$ above this region might be due to the instabilities that decelerate alpha particles as $V_A$
decreases with distance \citep{gary_electromagnetic_2000}, or due to Coulomb collisions that tend to bring all species to equilibrium \citep{marsch_solar_1982}.
It is also important to note that the amount of the sub-Alfv\'enic measurements is limited so far.
We expect more sub-Alfv\'enic observations for a more definite conclusion in the future.

\begin{figure}
    \centering
    \includegraphics[width=1.0\textwidth, height=0.75\textwidth]{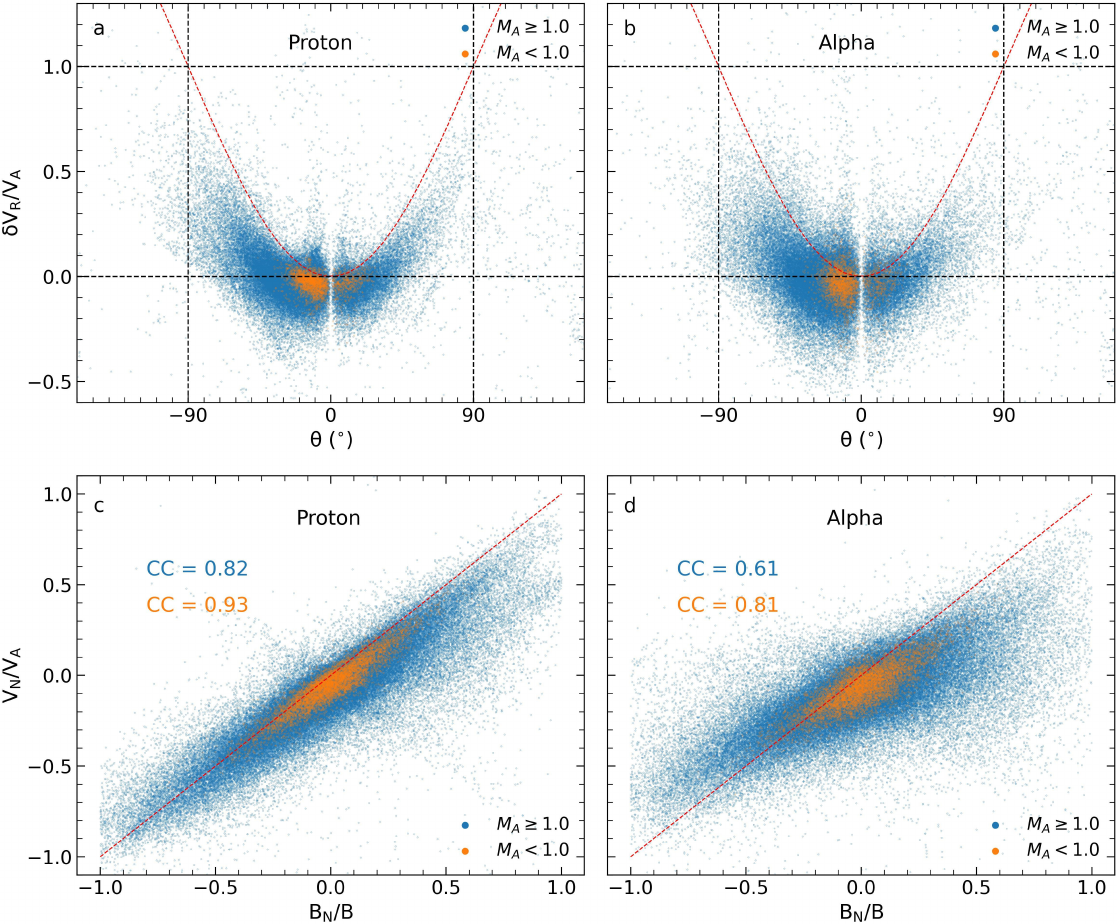}
    \caption{Effects of Alfvn\'eic fluctuations on protons (left) and alphas (right).
    Panels (a) and (b) show the radial velocity variations of protons and alphas in units of the local Alfv\'en speed as a function of magnetic field deflection angle.
    The red dashed curves are derived from the equation $\delta V_R / V_A = 1 - \mathrm{cos}\theta$ given by \cite{Liu_2023}.
    Panels (c) and (d) refer to the correlations between the normal components of the velocity of protons and alpha particles in units of the local Alfv\'en speed and the magnetic field.
    The red dashed lines indicate the slope corresponding to the local Alfv\'en speed (slope=1).
    The correlation coefficients (CCs) of the scatters are also given.
    Again, the orange scatter refers to the sub-Alfv\'enic measurement.}
    
    \label{fig:VN_VA_BN_B}
\end{figure}

For better understanding the effects of Alfv\'enic fluctuations across the CAS on the particles, 
we plot the behaviors of the two ion species in Figure \ref{fig:VN_VA_BN_B}.
Figure \ref{fig:VN_VA_BN_B}(a) displays the radial velocity enhancement of protons in units of the local Alfv\'en speed versus the magnetic field deflection angle.
Their relationship is consistent with the equation $\delta V_R / V_A = 1 - \mathrm{cos}\theta$ given by \cite{Liu_2023}.
The red dashed curve acts as an upper limit for the radial velocity enhancement because 
the magnetic field is not exactly radial or anti-radial.
Figure \ref{fig:VN_VA_BN_B}(a) confirms the findings of \cite{Liu_2023} but with larger statistics.
In the sub-Alfv\'enic solar wind, the trend is not very clear as the deflection angles are small.
The behaviors of alphas in the Alfv\'enic fluctuations, as shown in Figure \ref{fig:VN_VA_BN_B}(b), are different from those of protons,
and their velocity fluctuations do not quite follow the red curve.
This is consistent with the more two-sided velocity variations of alphas shown in Figure \ref{fig:en9} and Figure \ref{fig:en10}.
Therefore, alpha particles are less modulated by Alfv\'enic fluctuations than protons.
Figure \ref{fig:VN_VA_BN_B}(c) and (d) present the relation between $B_N / B$ and $V_N/V_A$ for protons and alphas, 
which can be used as another indicator of the effects of Alfv\'enic fluctuations \citep{goldstein_alfven_1995}.
Protons show a strong correlation between $B_N / B$ and $V_N/V_A$ close to the expected Alfv\'en wave relation (the red dashed line) for both below and above the CAS (correlation coefficient CC = $0.93$ and $0.82$, respectively).
A weaker correlation is seen for alphas in the super-Alfv\'enic solar wind (CC=$0.61$),
but in the sub-Alfv\'enic solar wind the correlation appears strong (CC=$0.81$).
It is surprising that below the CAS alpha particles are also considerably modulated by Alfv\'enic fluctuations.
Therefore, alphas may be significantly accelerated and heated by Alfv\'enic fluctuations below the CAS.
The increased correlation between $B_N / B$ and $V_N/V_A$ for alphas in the sub-Alfv\'enic solar wind 
may be due to $V_{\alpha p}$ being closer to 0 in this region.
According to \cite{mcmanus_density_2022}, the behavior of alphas in Alfv\'enic fluctuations is close to that of protons 
when $V_{\alpha p}$ is small.
The CC of alphas, however, is smaller than that of protons for both below and above the CAS.
\cite{matteini_ion_2015} 
analyzed the motions of protons and alphas in the Alfv\'en wave frame, 
and suggested that alphas are usually closer to the center of the wave frame, which may result in the weaker modulation of alphas by Alfv\'enic fluctuations. 
\cite{mcmanus_density_2022} demonstrated three scenarios, in which $V_N/V_A$ for alphas is correlated, uncorrelated, and anti-correlated with $B_N/B$.
The combination of these cases may lead to a decrease in the CC for alphas.

\section{Conclusions} \label{sec:con}
We have investigated the characteristics of the alpha-proton differential flow in the young solar wind across the CAS,
and the effects of Alfv\'enic fluctuations including switchbacks on the differential flow.
We summarize our findings, which address the three key questions that we have put forward in Section \ref{sec:intro}, as follows:
\begin{enumerate}
    \item In general, alphas stream faster than protons (in 90.3 \% of the cases), the differential velocity between them is 
          aligned with the local magnetic field vector, and the local Alfv\'en speed provides an upper limit for the differential velocity (in 97.9 \% of the cases).
          These results are consistent with previous studies. However, the situation changes with the solar wind speed.
          In the slowest solar wind ($V_p < 325$ km/s), we do observe occasions when alphas stream slower than protons (in 14.0 \% of the cases).
          As the solar wind becomes faster, the differential velocity in units of the local Alfv\'en speed also increases
          but remains smaller than 1.

    \item We find an overall enhancement of $V_{\alpha p}$ when $M_A \lesssim 2$, 
          which may indicate that a strong preferential acceleration of alphas occurs in this region.
          The preferential acceleration of alphas extends to higher $M_A$ in the faster solar wind.
          This may be attributed to the combination of two factors: the more intense and diverse preferential acceleration mechanisms
           and the less significant Coulomb collisions in the faster solar wind.
          After $M_A \simeq 2$, $V_{\alpha p}$ decreases with increasing $M_A$.
          Below the CAS, the differential velocity in units of the local Alfv\'en speed is generally smaller than 1 (in 99.9 \% of the cases), 
          and the case of $V_{\alpha} < V_{p}$ exists only in the slowest solar wind ($V_p < 175$ km/s).
          The situation is similar above the CAS. The differential velocity is generally smaller than the local Alfv\'en speed (in 97.8 \% of the cases),
          and $V_{\alpha p} < V_p$ cases are also seen only in the slowest solar wind ($V_p < 325$ km/s).

    \item Alfv\'enic fluctuations including switchbacks could affect the characteristics of the differential velocity.
          We observe deviations from the alignment between the differential velocity and the local magnetic field vector, and they are
          accompanied by large magnetic field deflection angles and drops in the differential velocity.
          Moreover, protons and alphas are modulated by Alfv\'enic fluctuations in different ways.
          The effects of Alfv\'enic fluctuations on protons are stronger than on alphas.
          These results are consistent with the findings of \cite{mcmanus_density_2022}.
          The radial velocity variations of alphas are more two-sided than those of protons.
          In the sub-Alfv\'enic solar wind, the correlation between $B_N/B$ and $V_N/V_A$ is strong for both protons and alphas (CC = 0.93 and 0.81, respectively).
          It is surprising that the effects of Alfv\'enic fluctuations on alphas appear strong in the sub-Alfvn\'eic solar wind.
          In the super-Alfv\'enic solar wind, the correlation between $B_N/B$ and $V_N/V_A$ is still strong for protons (CC = 0.82),
          but for alphas the correlation is weakened (CC=0.61).

    \item Apart from the above conclusions that correspond to the three key questions in Section \ref{sec:intro},
          we also have some extra findings.
          We reveal that the deflection angle increases with the solar wind speed in the super-Alfv\'enic solar wind, 
          while in the sub-Alfv\'enic solar wind the deflection angle is concentrated at small values in all speed categories.
          We also find that the preference for clockwise deflections is observed not only in the super-Alfv\'enic wind, but also in the sub-Alfv\'enic wind.
          The degree of the preference decreases as the solar wind speed increases, which may result from the more radial background Parker spiral field in the faster solar wind.
        
\end{enumerate}

\section*{Appendix: Data Selection and Curation} \label{sec:app}
\begin{figure}
    \centering
    \includegraphics[width=0.90\textwidth, height=1.0\textwidth]{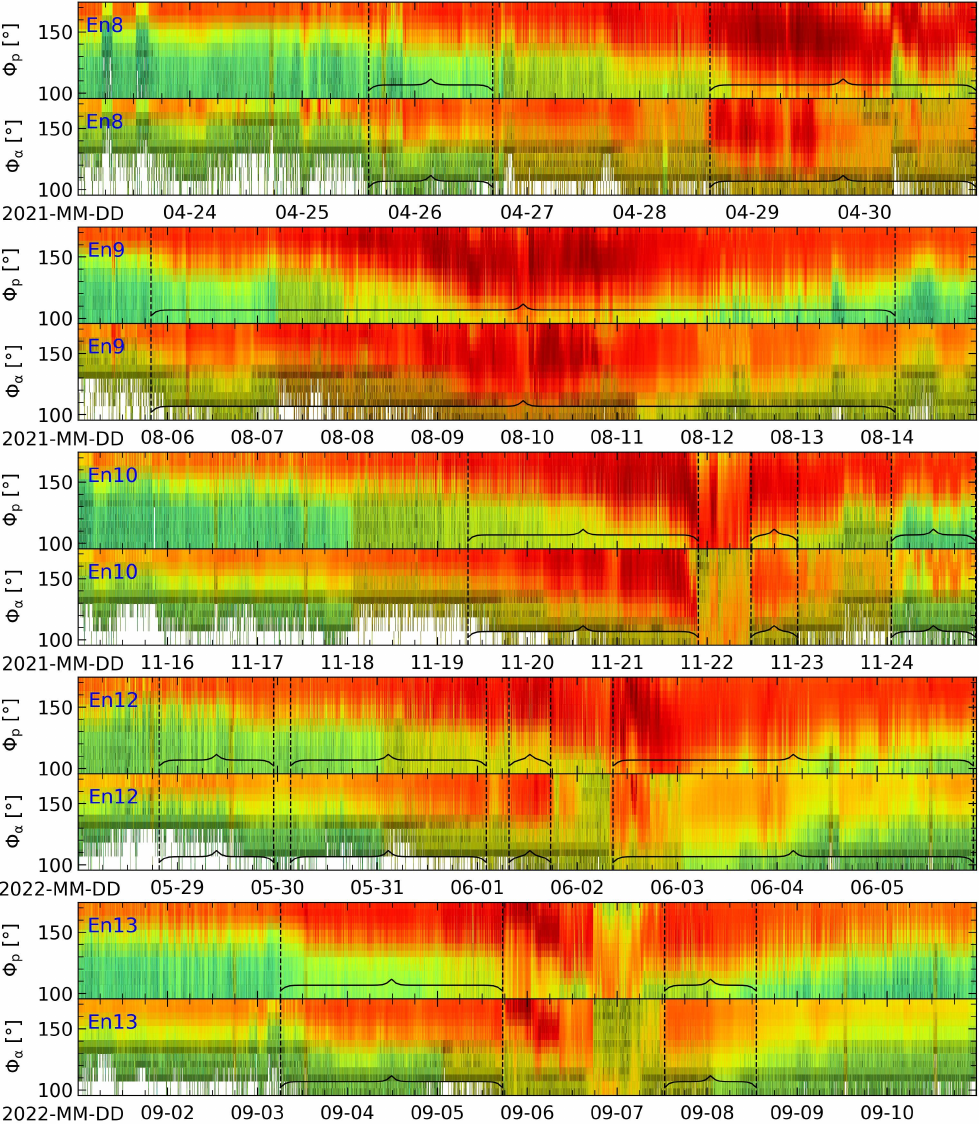}
    \caption{The azimuthal fluxes of protons (upper) and alphas (lower) in the spacecraft frame over time for each encounter.
    Dashed lines and curly brackets indicate the intervals that are considered reliable.}
    
    \label{fig:selection}
\end{figure}

In order to have reliable velocity measurements, 
we need to select intervals when the cores of the solar wind ions (alphas and protons) are primarily within 
the field of view (FOV) and energy range of SPAN-I.
For each encounter, the total length of the reliable intervals is usually a few days.
Specifically, our selection is derived through the following steps.
We select intervals in which the peak of ion flux is in the energy range of SPAN-I. 
The bulk of the 3D velocity distribution functions (VDFs) of the ions is also confirmed to be within the FOV of SPAN-I during these intervals.
Figure \ref{fig:selection} shows the azimuthal fluxes of alphas and protons in the spacecraft frame.
The selected time intervals are indicated by the dashed lines and curly brackets in Figure \ref{fig:selection}.
Note that the CME at encounter 13 is already removed from the data in this step.
Readers are redirected to \cite{liu2024CME} for the time interval of the CME.
Table 1 lists the detailed time ranges of the selected intervals and the HCS crossings within them.
The relatively long duration (up to several hours) of the HCS crossings, as shown 
in Table 1, is a result of their clustered occurrence.
Our final dataset is obtained by subtracting the times of the clustered HCS crossings from the selected intervals.

\section*{Acknowledgments}
This research was supported by NSFC under grant 42274201,
the National Key R\&D Program of China (No.2021YFA0718600 and No.2022YFF0503800),
the Strategic Priority Research Program of the Chinese Academy of Sciences (No.XDB0560000),
and the Specialized Research Fund for State Key Laboratories of China.
C.C. was supported by the Research Foundation of Education Bureau of Hunan Province, China (No.23B0593).
P.M. acknowledges the support from NASA HGIO grant 80NSSC23K0419.
We are grateful to Dr. Huidong Hu and Dr. Rui Wang for helpful discussions.
We thank the NASA PSP mission and the SWEAP and FIELDS teams for allowing uses of their data.
The FIELDS RTN magnetic field data and the QTN data are available at \href{https://research.ssl.berkeley.edu/data/psp/data/sci/fields/}{https://research.ssl.berkeley.edu/data/psp/data/sci/fields/},
and the SWEAP ion data are available at \href{http://sweap.cfa.harvard.edu/pub/data/sci/sweap/spi/L3/}{http://sweap.cfa.harvard.edu/pub/data/sci/sweap/spi}.
We acknowledge the work of the \textit{Sunpy} team \citep{sunpy_community2020}.

\bibliography{sample631}{}
\bibliographystyle{aasjournal}

\clearpage

\begin{table}[!ht]
    \begin{center}
        \renewcommand\arraystretch{0.7}
        \begin{tabular}{p {3cm} | p {3cm} | p {3cm} | p {3cm} | p {3cm}}
            \hline
            \hline
            \multirow{2}{*}{Encounter} & \multicolumn{2}{c|}{Selected time intervals} & \multicolumn{2}{c}{Clustered HCS crossings} \\ \cline{2-5}
            \multirow{2}{*}{ } & Star time (UT) & End time (UT) & Start time (UT) & End time (UT) \\
            \hline
            \multirow{2}{*}{8} & 2021-04-25 14:10 & 2021-04-26 16:40 & 2021-04-29 00:30 & 2021-04-29 02:10 \\ \cline{2-5}
            \multirow{2}{*}{ } & 2021-04-28 15:00 & 2021-04-30 23:50 & 2021-04-29 08:00 & 2021-04-29 14:45 \\ 
            \hline
            \multirow{2}{*}{9} & \multirow{2}{*}{2021-08-05 19:40} & \multirow{2}{*}{2021-08-14 02:10} & 2021-08-10 00:10 & 2021-08-10 02:20 \\ \cline{4-5}
            \multirow{2}{*}{ } & \multirow{2}{*}{ } & \multirow{2}{*}{ } & 2021-08-10 09:00 & 2021-08-10 18:20 \\ \cline{4-5}
            \hline   
            \multirow{3}{*}{10} & 2021-11-19 08:10 & 2021-11-21 21:40 & \multirow{3}{*}{Not Found} & \multirow{1}{*}{} \\ \cline{2-3}
            \multirow{3}{*}{ } & 2021-11-22 11:40 & 2021-11-23 00:10 & \multirow{3}{*}{} & \multirow{1}{*}{} \\ \cline{2-3}
            \multirow{3}{*}{ } & 2021-11-24 01:10 & 2021-11-24 23:40 & \multirow{3}{*}{} & \multirow{1}{*}{}  \\ \cline{2-3}
            \hline
            \multirow{4}{*}{12} & 2022-05-28 19:40 & 2022-05-29 23:10 & \multirow{4}{*}{2022-06-02 11:30} & \multirow{4}{*}{2022-06-02 20:30} \\ \cline{2-3}
            \multirow{4}{*}{ } & 2022-05-30 03:10 & 2022-06-01 02:10 & \multirow{4}{*}{ } & \multirow{4}{*}{ } \\ \cline{2-3}
            \multirow{4}{*}{ } & 2022-06-01 07:40 & 2022-06-01 17:40 & \multirow{4}{*}{ } & \multirow{4}{*}{ } \\ \cline{2-3}
            \multirow{4}{*}{ } & 2022-06-02 08:40 & 2022-06-05 23:10 & \multirow{4}{*}{ } & \multirow{4}{*}{ } \\ \cline{2-3}
            \hline
            \multirow{2}{*}{13} & 2022-09-03 06:10 & 2022-09-05 17:30 & \multirow{2}{*}{Not Found} & \multirow{2}{*}{ } \\ \cline{2-3}
            \multirow{2}{*}{ } & 2022-09-07 12:40 & 2022-09-08 13:10  & \multirow{2}{*}{ } & \multirow{2}{*}{ } \\ \cline{2-3}
            \hline
        \end{tabular}
        \caption{The selected reliable time intervals and clustered HCS crossings for each encounter.}
        \label{table:selection}
    \end{center}
    \end{table}




\end{document}